\documentclass[12pt,thmsa]{article}
\usepackage{sw20lart}

\input{tcilatex}
\begin{document}

\title{On possible memory effects in tests of Bell inequalities }
\author{Emilio Santos \\
Departamento de F\'{i}sica, Universidad de Cantabria. Santander. Spain}
\date{March, 21, 2016 }
\maketitle

\begin{abstract}
It is shown that memory effects in experiments measuring correlations in
entangled photon pairs are not able to produce a relevant loophole for the
test of local hidden variables theories.
\end{abstract}

\section{Introduction}

The Bell inequalities\cite{Bell} are necessary conditions for the existence
of local realistic (or local hidden variables) models able to interpret the
results of some specific experiments. In particular the experiments
involving coincidence measurements made by two separated parties, Alice and
Bob, on entangled pairs of particles. The relevance of Bell's work is that
there are idealized experiments where quantum mechanics predicts violations
of the inequalities, which strongly suggests that for some actual
experiments no local hidden variables model is possible. The contradiction
between local realism and quantum mechanics is known as ``Bell's theorem''.
The question whether the contradiction also exists for actual, rather than
ideal, experiments has led people to perform many Bell tests. Until recently
all tests have provided results compatible with both quantum mechanics and
local realistic models, the latter compatibility deriving from the existence
of loopholes in the tests. Last year several experiments have been performed
that, for the first time, blocked simultaneously the two most relevant
loopholes, detection inefficiency and locality\cite{Hensen}, \cite{Shalm}, 
\cite{Giustina}. The purpose of this note is to see whether a loophole still
remains, namely the \textit{memory loophole}.

The memory loophole derives from the fact that tests of the Bell
inequalities require performing many similar experiments on different\textit{%
\ }entangled pairs of particles in order to estimate the probabilities from
the frequencies actually measured. In practice there is a series of trials
with a given experimental set-up, the trials being separated from each other
by short time intervals. It is usually assumed that any hidden variables
associated with the nth particle pair would be independent of measurement
choices and outcomes for the first (n-1) pairs. Models which violate this
assumption exploit the possible memory effects. The strongest type of
violation uses a 2-sided memory loophole, in which the hidden variables for
pair n can depend on the previous measurement choices and outcomes in both
wings of the experiment.

The memory loophole has been studied by Barrett et al.\cite{Barrett}. The
authors concluded that, although in principle the memory loophole implies a
slight flaw in existing analyses of Bell experiments, the data still may
refute local realistic models. However Barrett et al. proved the irrelevance
of the memory loophole for tests of the CHSH (Bell type) inequality\cite
{CHSH}, whilst the recent photon experiments test the CH-Eberhard inequality%
\cite{Eberhard}. The point is that the CHSH\ inequality $\left( \ref{CHSH}%
\right) $ is apropriate for event-ready detectors, like in the experiment by
Hensen et al.\cite{Hensen}, where the probabilities may be calculated on a
well defined set of entangled particle pairs. In contrast the CH-Eberhard
inequality ($\ref{p}$) is used when the set of pairs is not well defined, as
in the recent photon experiments. In these experiments the photon pairs are
produced via parametric down conversion in a nonlinear crystal. The
consequence is that when neither Alice nor Bob detect a photon in a trial
they cannot know if either photons arrived to the parties but none was
detected or no photon was produced in the source during the trial. In fact
the second case is by far more probable. Thus in the photon experiments we
cannot determine true probabilities but only relative probabilities. The
question arises whether the memory loophole is also irrelevant in these
conditions. In the following I give an affirmative answer to the question.

In photon experiments in more detail the Bell test is divided into a series
of trials. During each trial Alice and Bob randomly choose between one of
two measurement settings, denoted $a$ and $a^{\prime }$ for Alice and $b$
and $b^{\prime }$ for Bob, and record either a ``$+$'' if they observe any
detection event or a ``$0$'' otherwise. The details of that type of Bell
test may be seen, for instance, in the experiments by Shalm et al.\cite
{Shalm} or by Giustina et al.\cite{Giustina}. Alice may get two possible
results in each one of the two possible measurements of her photon, and
similar for Bob. Therefore there are 16 coincidence probabilities that might
be determined. In practice the Bell test requires just 4 that may be
combined in the CH/Eberhard inequality\cite{Eberhard}, namely 
\begin{eqnarray}
0 &\leq &B\equiv P\left( 0+\mid ab^{\prime }\right) +P\left( +0\mid
a^{\prime }b\right)   \label{p} \\
&&+P\left( ++\mid a^{\prime }b^{\prime }\right) -P\left( ++\mid ab\right) . 
\nonumber
\end{eqnarray}
The terms $P(++\mid ab)$ and $P(++\mid a^{\prime }b^{\prime })$ correspond
to the probability that both Alice and Bob record detection events $(++)$
when they choose the measurement settings $ab$ or $a^{\prime }b^{\prime }$,
respectively. Similarly, the terms $P(+0\mid ab^{\prime })$ and $P(0+\mid
a^{\prime }b)$ are the probabilities that only Alice or Bob record an event
for settings $ab^{\prime }$ and $a^{\prime }b$, respectively.

For comparison I write the CHSH inequality, that is 
\begin{equation}
E(ab)+E(a^{\prime }b)+E(ab^{\prime })-E(a^{\prime }b^{\prime })\leq 2,
\label{CHSH}
\end{equation}
where $E(ab)$ denotes the expectation value of the product of the outcomes
of the measurements with the settings $a$ and $b$. With the notation of eq.$%
\left( \ref{p}\right) $ we have 
\[
E(ab)=P\left( ++\mid ab\right) +P\left( 00\mid ab\right) -P\left( +0\mid
ab\right) -P\left( 0+\mid ab\right) .
\]
Actually the inequalities $\left( \ref{p}\right) $ and $\left( \ref{CHSH}%
\right) $ are equivalent, in the sense that one of them is fulfilled if and
only if the other one holds true, provided that the label $``0"$ has the
same meaning in both. But this is not obviously the case when we compare
event ready experiments, e. g. by Hensen et al.\cite{Hensen} with
experiments with photons produced via parametric down conversion, like those
of Giustina et al.\cite{Giustina} or Shalm et al.\cite{Shalm}. Indeed in the
latter type of experiment the CHSH inequality is useless because most of the
times no photon arrives at the detectors, whence we would have 
\[
E(ab)\simeq P\left( 00\mid ab\right) \simeq 1.
\]
As a result the left side of eq.$\left( \ref{CHSH}\right) $ is very close to
2 and, what is more relevant, it cannot be calculated precisely because the
fraction of trials corresponding to no photons emitted from the source is
uncertain.

In spite of these difficulties it is possible to prove that the test of
local hidden variables is reliable. The proof follows. The joint probability 
$P(x,y;a,b)$ that the particles yield the outcomes $x$ and $y$ when
subjected to the measurement with the settings $a$ and $b$ respectively is
given, according to Bell\cite{Bell}, by 
\begin{equation}
P(x,y;a,b)=\int d\lambda \rho (\lambda )P(x;a,\lambda )P(y;b,\lambda ),
\label{Bell}
\end{equation}
where, in our notation, $x,y$ might be either $+$ or $0$. Eq.$\left( \ref
{Bell}\right) $ may be taken as the definition of local hidden variables
model. Now let us assume that there are memory effects such that the hidden
variables, $\lambda ,$ and the action of the measuring devices in the trial $%
n$ depend on all previous trials, $j=1,2,...n-1.$ This means that we shall
substitute the following

\begin{equation}
P_{n}(x,y;a,b)=\int d\lambda _{n}\rho _{n}(\lambda _{n})P_{n}(x;a,\lambda
_{n})P_{n}(y;b,\lambda _{n}),  \label{mem}
\end{equation}
for eq.$\left( \ref{Bell}\right) .$ That is, all possible memory effect may
just change the nature and the probability distribution of the hidden
variables, $\lambda ,$ and the probability of outcome in the measurement for
given settings, $a$ and $b$. If the probabilities of the 4 outcomes $%
ab,ab^{\prime },a^{\prime }b,a^{\prime }b^{\prime }$ are 1/4, as insured by
the random choice, then inserting eq.$\left( \ref{memory}\right) $ in eq.$%
\left( \ref{p}\right) $ we get the CH-Eberhard inequality 
\begin{eqnarray}
0 &\leq &B_{n}\equiv P_{n}\left( 0+\mid ab^{\prime }\right) +P_{n}\left(
+0\mid a^{\prime }b\right)   \label{pn} \\
&&+P_{n}\left( ++\mid a^{\prime }b^{\prime }\right) -P_{n}\left( ++\mid
ab\right) .  \nonumber
\end{eqnarray}
This happens if an entangled pair of photons is produced in the source in
the nth trial, but if no photons are produced then eq.$\left( \ref{p}\right) 
$ gives 
\[
B_{n}=0.
\]
In any case for a set of trials we shall have 
\[
\sum_{j}B_{j}\geq 0,
\]
where $\left\{ j\right\} $ represents the trials chosen for the test of the
CH-Eberhard inequality. In practice it is common to choose, within some time
interval, all trials  such that at least one photon is detected (either by
Alice or by Bob). The absence of bias in the random choice of the settings
is essential for the proof. For instance the local model 
\[
P(+;a,\lambda )=P(+;b,\lambda )=P(+;a^{\prime },\lambda )=1,P(+;b^{\prime
},\lambda )=0,
\]
gives $B=0$, taking eqs.$\left( \ref{Bell}\right) $ and $\left( \ref{p}%
\right) $ into account, for unbiased random choices. However if the choices $%
a$ and $b$ have probabilities $(1+\varepsilon )/2$ each and the choices $%
a^{\prime }$ and $b^{\prime }$ probabilities $(1-\varepsilon )/2,$ then we
get $B=-\varepsilon $ in apparent violation of eq.$\left( \ref{p}\right) .$

We conclude that for photon tests involving the CH-Eberhard inequality the
possible memory loophole is irrelevant provided that the test involves a
large enough number of trials. For a small number there may be fluctuations
that could give a wrong answer. The study of fluctuations will not be made
here, it would be similar to the study made by Barrett et al.\cite{Barrett}
in relation with the CHSH inequality.

\end{document}